\documentclass[namedreferences,hyperref,optionalrh,solaromanenum]{spr-sola}

\usepackage{graphicx} 
\usepackage{color} 

\ifx\urlurl \undefined
\def\urlurl#1{\href{http://#1}{\textsf{#1}}}\fi

\chardef\us=`\_

\begin{document}

\begin{frontmatter}

\title{A Study on Magnetic-sensitivity Wavelength Position of the Working Line Used by the {\it Full-Disk Magnetograph} onboard the {\it Advanced Space based Solar Observatory} (ASO-S/FMG)}

%
\author[addressref={aff1,aff2,aff3},corref,email={lius@nao.cas.cn}]{\inits{S.}\fnm{S.}~\lnm{Liu} \orcid{0000-0002-1396-7603}}
\author[addressref={aff1,aff2,aff3},email={sjt@nao.cas.cn}]{\inits{J.T.}\fnm{J.T.}~\lnm{Su}\orcid{0000-0002-5152-7318}}
\author[addressref={aff1,aff2,aff3}]{\inits{X.Y.}\fnm{X.Y.}~\lnm{Bai}\orcid{0000-0003-2686-9153}}
\author[addressref={aff1,aff2,aff3}]{\inits{Y.Y.}\fnm{Y.Y.}~\lnm{Deng}\orcid{0000-0003-1988-4574}}
\author[addressref={aff1,aff2,aff3}]{\inits{J.}\fnm{J.}~\lnm{Chen}\orcid{0000-0001-7472-5539}}

\author[addressref={aff1,aff2,aff3}]{\inits{Y.L.}\fnm{Y.L.}~\lnm{Song} \orcid{0000-0002-9961-4357}}

\author[addressref={aff1,aff2,aff3}]{\inits{X.F.}\fnm{X.F.}~\lnm{Wang} \orcid{0000-0003-2776-3895}}

\author[addressref={aff1,aff2,aff3}]{\inits{H.Q.}\fnm{H.Q.}~\lnm{Xu}\orcid{0000-0003-4244-1077}}

\author[addressref={aff1,aff2,aff3}]{\inits{X.}\fnm{X.}~\lnm{Yang} \orcid{0000-0003-1675-1995}}
\author[addressref={aff3,aff1,aff2}]{\inits{S}\fnm{Shahid}~\snm{Idrees}\orcid{0009-0008-2912-1136}}
%

\runningauthor{S. Liu et al.}
\runningtitle{The Spectral Sensitivity of ASO-S/FMG}

\address[id=aff1]{National Astronomical Observatories, Chinese Academy of Science, Beijing, 100101, China}
\address[id=aff2]{Key Laboratory of Solar Activity and Space Weather, National Space Science Center, Chinese Academy of Science, Beijing, 100190, China}
\address[id=aff3]{School of Astronomy and Space Sciences, University of Chinese Academy of Sciences, Beijing, 100101, China}
\begin{abstract}
Utilizing data from the $Solar$ $Magnetism$ and $Activity$ $Telescope$ (SMAT), analytical solutions of polarized radiative transfer equations, and in-orbit test data from the Full-disk Magnetograph (FMG) onboard the \textit{Advanced Space based Solar Observatory} (ASO-S), this study reveals the magnetic-sensitivity spectral positions for the Fe {\sc i} $\lambda$5234.19\,\AA{} working line used by FMG. From the experimental data of SMAT, it is found that the most sensitivity position is located at the line center for linear polarization (Stokes-Q/U), while it is about -0.07\,\AA~away from the line center for circular polarization (Stokes-V).  Moreover, both the
theoretical analysis and the in-orbit test data analysis of FMG prove again the above results. Additionally, the theoretical analysis suggests the presence of distinct spectral pockets (centered at 0.08-0.15\,\AA) from the line, harboring intense magnetic sensitivity across all three Stokes parameters. Striking a balance between high sensitivity for both linear and circular polarization while capturing additional valuable information, a spectral position of -0.08\,\AA~emerges as the champion for routine FMG magnetic-field observations.

\end{abstract}

\end{frontmatter}

\section{Introduction}\label{S-Introduction} 
Measuring the solar magnetic field on the photosphere typically involves analyzing the inversion of polarized spectra based on the Zeeman effect. Two main instrument types serve this purpose: One of them is known as polarization spectrographs, used to measure the Stokes parameter profiles of magnetically sensitive lines. Subsequently, a specific atmospheric model is used to invert these profiles and derive the magnetic field.
Examples include the Advanced Stokes Polarimeterat the \textit{Vacuum Tower Telescope} (VTT/ASP: \citeauthor{1992SPIE.1746...22E} \citeyear{1992SPIE.1746...22E}) of the National Solar Observatory, and the Spectro-Polarimeter of the \textit{Solar Optical Telescope} (SOT/SP: Tsuneta et al. 2008) onboard Hinode.
Other one is called vector video magnetographs (filter instruments), employ limited spectral sampling to capture Stokes parameter images. Calibration or inversion techniques then translate these images into magnetic field information.
Examples include the \textit{Solar Magnetic Field Telescope} at Huairou Solar Observing Station (HSOS/SMFT: \citeauthor{1986AcASn..27..173A} \citeyear{1986AcASn..27..173A}), the \textit{Michelson Doppler Imager} on board the Solar and Heliospheric Observatory (SOHO/MDI: \citeauthor{1995SoPh..162..129S}~\citeyear{1995SoPh..162..129S}), and the \textit{Helioseismic and Magnetic Imager} on board the Solar Dynamical Observatory (SDO/HMI: \citeauthor{2012SoPh..275..207S} \citeyear{2012SoPh..275..207S}; \citeauthor{2012SoPh..275..229S} \citeyear{2012SoPh..275..229S}).
While high spectral resolution grants polarization spectrographs superior magnetic field measurement accuracy compared to vector magnetographs, they sacrifice temporal resolution. However, increasing the spectral sampling of vector magnetographs can enhance the agreement between the two instrument types (\citeauthor{1985svmf.nasa..342L} \citeyear{1985svmf.nasa..342L}; \citeauthor{1999SoPh..189....1L} \citeyear{1999SoPh..189....1L}).

For filter-type instruments utilizing single spectral sampling, choosing a magnetically sensitive position within the working line is crucial. The increased sensitivity translates directly to more reliable magnetic field measurements. Hence, investigations into spectral line positions with optimal magnetic sensitivity are of paramount importance.
The Full-disk vector magnetograph onboard tge \textit{Advanced Space based Solar Observatory} (ASO-S/FMG) is dedicated to full-disk photospheric magnetic field observations (\citeauthor{2019RAA....19..157D} \citeyear{2019RAA....19..157D}; \citeauthor{2019RAA....19..156G}, \citeyear{2019RAA....19..156G}; \citeauthor{2019RAA....19..161S} \citeyear{2019RAA....19..161S}).
In its routine mode, it acquires polarization images at a single spectral position. Therefore, determining the magnetically sensitive position of the working line is critical for FMG. Its chosen line, Fe {\sc i} $\lambda$5234.19 \AA{}, exhibits strong magnetic sensitivity due to its Land{\'e} factor of 1.5 (g=1.5) and total half-width of 0.334\,\AA{} (\citeauthor{2019RAA....19..157D} \citeyear{2019RAA....19..157D}; \citeauthor{2019RAA....19..161S} \citeyear{2019RAA....19..161S}; \citeauthor{2020ChPhB..29l4211H} \citeyear{2020ChPhB..29l4211H}; \citeauthor{2023SoPh..298...68G}, \citeyear{2023SoPh..298...68G}).

This study embarks on a comprehensive exploration of the magnetic sensitivity within the working line utilized by the FMG. We leverage three distinct vantage points; (i) As our experimental data, we draw upon photospheric magnetic field observations acquired by the \textit{Solar Magnetism and Activity Telescope} (SMAT) at the Huairou Solar Observing Station (\citeauthor{2007ChJAA...7..281Z} \citeyear{2007ChJAA...7..281Z}). 
(ii) We delve into the analytical solutions of radiative transfer equations, investigating the resulting response functions and their relationship to magnetic field strength (\citeauthor{1982SoPh...78..355L} \citeyear{1982SoPh...78..355L}; \citeauthor{1994A&A...283..129R} \citeyear{1994A&A...283..129R}; \citeauthor{2007A&A...462.1137O} \citeyear{2007A&A...462.1137O}). (iii) Finally, we analyze in-orbit test data from FMG itself, offering an additional perspective on our experimental and theoretical findings. This valuable insight also informs the optimal operation of FMG.

The structures of this article are arranged as follows: Section~\ref{sect:Experimental Test} details the descriptions and experimental results obtained with SMAT.
Section~\ref{sec:Theoretical Analysis} delves into the theoretical analysis of radiative transfer equations and their corresponding response functions to varying magnetic field strengths.
Section~\ref{sect:The FMG Test} dissects the analysis of FMG's in-orbit test data. Finally, Section~\ref{sec:disc-conc} provides a comprehensive discussion and conclusive remarks stemming from the combined knowledge gleaned from each data source.

\section{Analysis of the Experimental Data of SMAT}
\label{sect:Experimental Test}
In 2006, the \textit{Solar Magnetism and Activity Telescope} (SMAT), with its 10 cm aperture and effective field of view of 32$^{'}$$\times$32$^{'}$, embarked on a mission to map the full-disk magnetic field of the solar photosphere.
Equipped with a 0.125\,\AA~bandpass birefringent filter, SMAT operates in two distinct modes.
First one know as routine observation mode in which the magnetograph focuses on a single pre-determined wavelength within the working line, capturing Stokes-Q, U, and V images. Second one called scanning mode which unlocks a wider spectral range, scanning from $-1$ \AA{} in the blue wing to $+1$ \AA{} in the red wing relative to the working line center. This broader perspective allows for the construction of full-disk magnetograms using a 992$\times$992 pixels CCD camera. Our experiment leveraged SMAT's scanning mode on 07 September 2017, acquiring 93 datasets (encompassing 31 individual sets for each of Q, U, and V). The scan meticulously covered a range of -0.3\,\AA~to 0.3\ around the line center with a step size of 0.02\,\AA, taking approximately 30 minutes to complete.

\begin{figure}
\centerline{\includegraphics[width=1\textwidth,clip=]{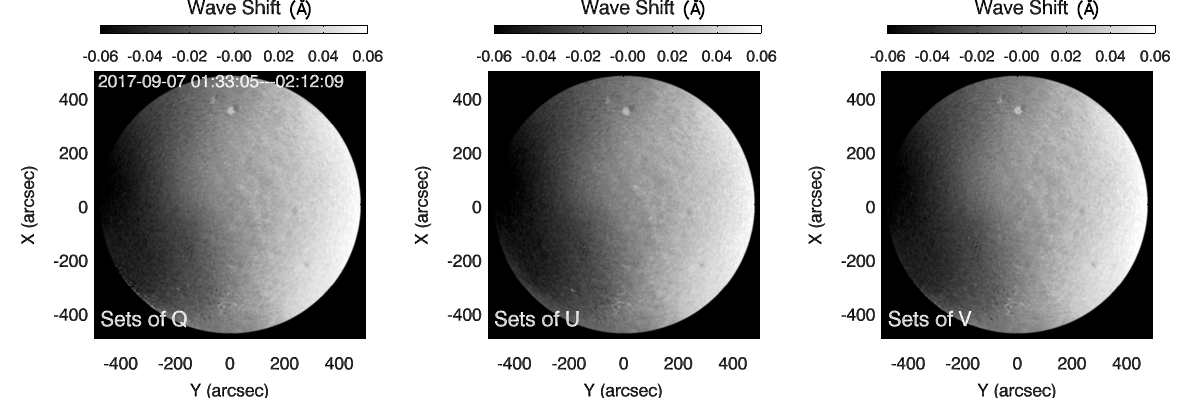}}
\caption{Wavelength shifts in unit of \AA~relative to that of heliocentric center obtained for three sets of Q, U and V observations (I$\pm$Q, I$\pm$U, and I$\pm$V).
} \label{smat_lqu_lincen_disk}
\end{figure}

Figure \ref{smat_lqu_lincen_disk} visually portrays the full-disk distributions of the measured wavelength shifts relative to the heliocentric center. These shifts were calculated by: (i) Fitting the line centers of intensity profiles at various positions across the solar disk. (ii) Subtracting the line center values at each position from the value at the disk center. Encouragingly, the wavelength shifts obtained from the Q, U, and V profiles display remarkable consistency, further highlighting the telltale signature of solar rotation in the data.

\begin{figure}

\centerline{\includegraphics[width=1\textwidth,clip=]{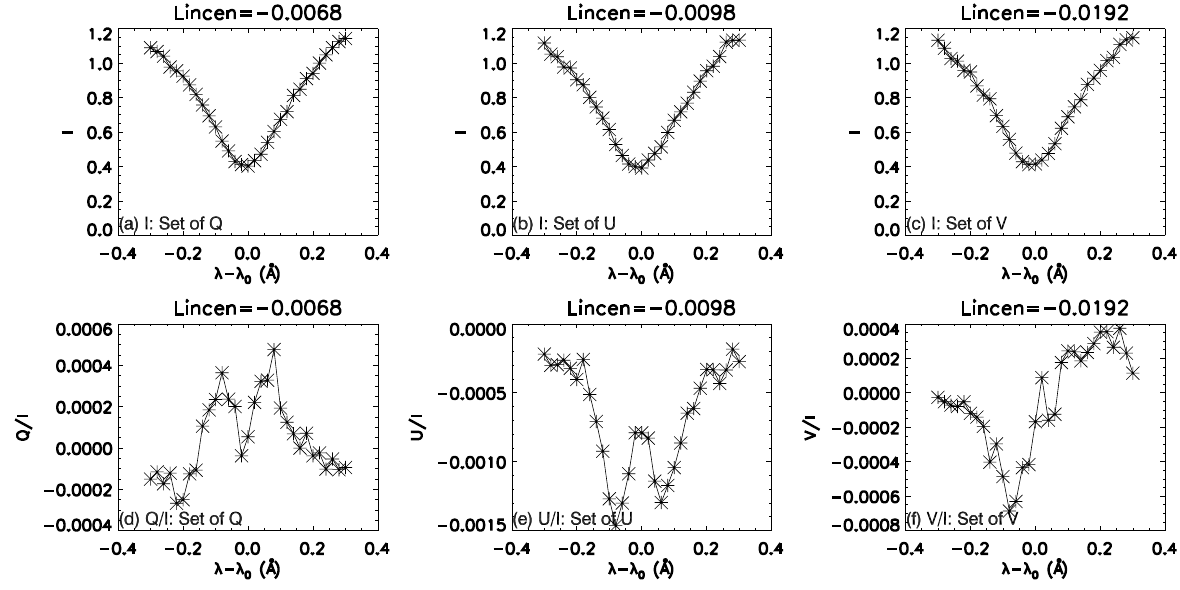}}
\caption{Stokes I, Q, U and V profiles in an active region, in which $x$-axis indicates the wavelength offset relative to line center (range: -0.3~\AA~to 0.3~\AA). Panels \textbf{(a)}, \textbf{(b)}, and \textbf{(c)}: Non-polarization intensity (I) profiles obtained from three separate sets of Q, U, and V observations. The respective line centers are calculated and highlighted within each panel. Panels \textbf{(d)}, \textbf{(e)} and \textbf{(f)}: Normalized profiles of (d) Q/I, (e) U/I, and (f) V/I. These ratios reveal fundamental characteristics of the spectral line, with V exhibiting an antisymmetric profile and Q and U demonstrating symmetric profiles.} \label{quvprofile}
\end{figure}

Figure \ref{quvprofile} showcases the Q, U, and V profiles extracted from an active region exhibiting strong polarization. The horizontal axis denotes the relative position of the spectral line to its center, labeled as $\lambda-\lambda_{0}$ in unit of \AA. The top panels (a, b, and c) depict the non-polarization intensity (I) profiles for each polarization state (Q, U, and V). Notably, the line centers are marked within each panel. Below, the bottom panels (d, e, and f) showcase the profiles of Q/I, U/I, and V/I, respectively. These normalized profiles reveal fundamental characteristics: V exhibits an antisymmetric profile, while Q and U demonstrate symmetric profiles. Figure \ref{quvprofile} serves as a reliable reference for our analysis due to its representation of typical spectral line profiles.


\begin{figure}
\centerline{\includegraphics[width=1\textwidth,clip=]{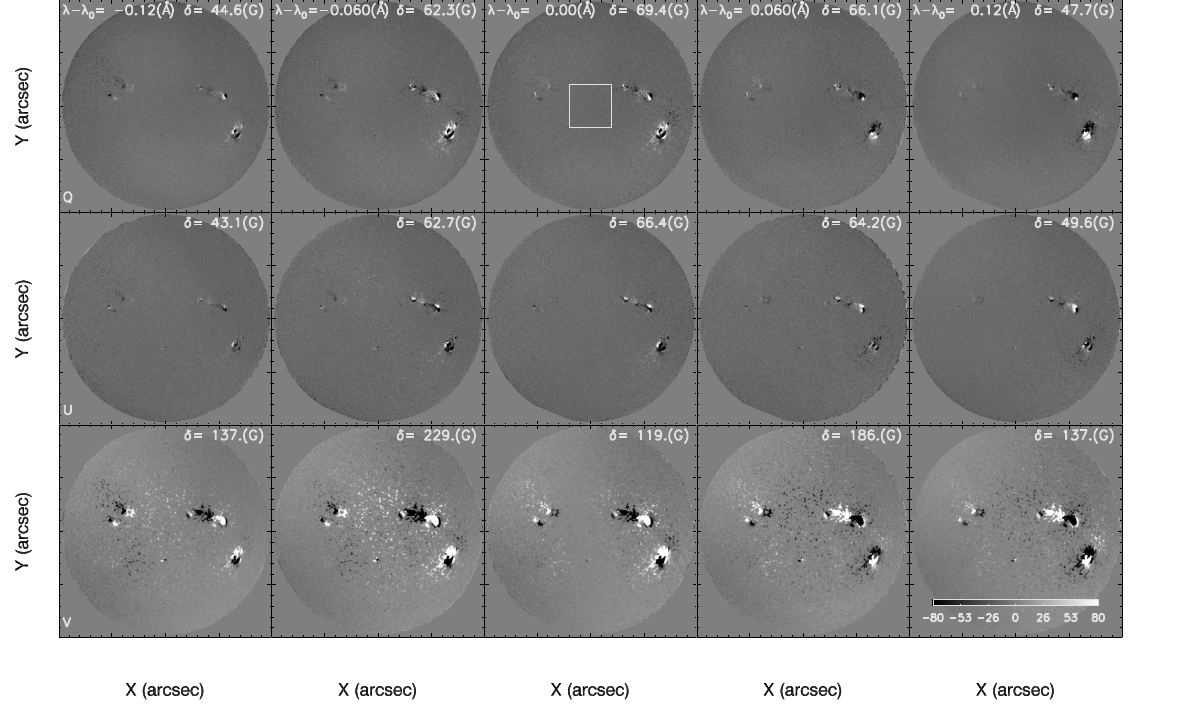}}
\caption{Top row: Stokes-Q images captured on 07 September 2017 at different wavelength positions relative to the line center ($\lambda-\lambda_{0}$ in \AA). Each panel displays the variance ($\delta$) of the magnetic field within a quiet region (outlined by a white square) at its corresponding wavelength. Similar images for Stokes-U and V are presented in the middle and bottom rows, respectively.} \label{31q}
\end{figure}

Moving on to Figure \ref{31q}, the top row displays full-disk Stokes-Q images captured at different wavelengths. The image at the line center occupies the central position, with the relative wavelength deviations ($\lambda-\lambda_{0}$) labeled above each panel. Examining these images, it appears that the line center exhibits the highest magnetic sensitivity. However, for a more precise evaluation, we calculate the variance ($\delta$) of the Stokes images within each image. We define $\delta$ as: $\delta_x$=1/N-1*$\sum_{i=0}^{N-1} \sqrt{(\bar{x}-x_i})^2$ for a series of values $x_i$ as signals, where $\bar{x}$ is the average of $x_i$, and N is the number of pixels. Generally, the larger $\delta_x$ is and the stronger signal is. To quantify this, we calculate the $\delta_x$ values for Q, U, and V within a quiet region at the disk center (e.g., indicated by a white square in the middle column of top row), marking them on each panel. Interestingly, for Stokes-Q measurements, the peak sensitivity occurs around the line center with a $\delta_x$ value of 69.4 G. Similarly, the middle and bottom rows reveal analogous results for Stokes-U and V. While U also exhibits peak sensitivity at the line center, V demonstrates its highest sensitivity slightly blueward of the line center, at approximately -0.06\,\AA.

\begin{figure}
\centerline{\includegraphics[width=1\textwidth,clip=]{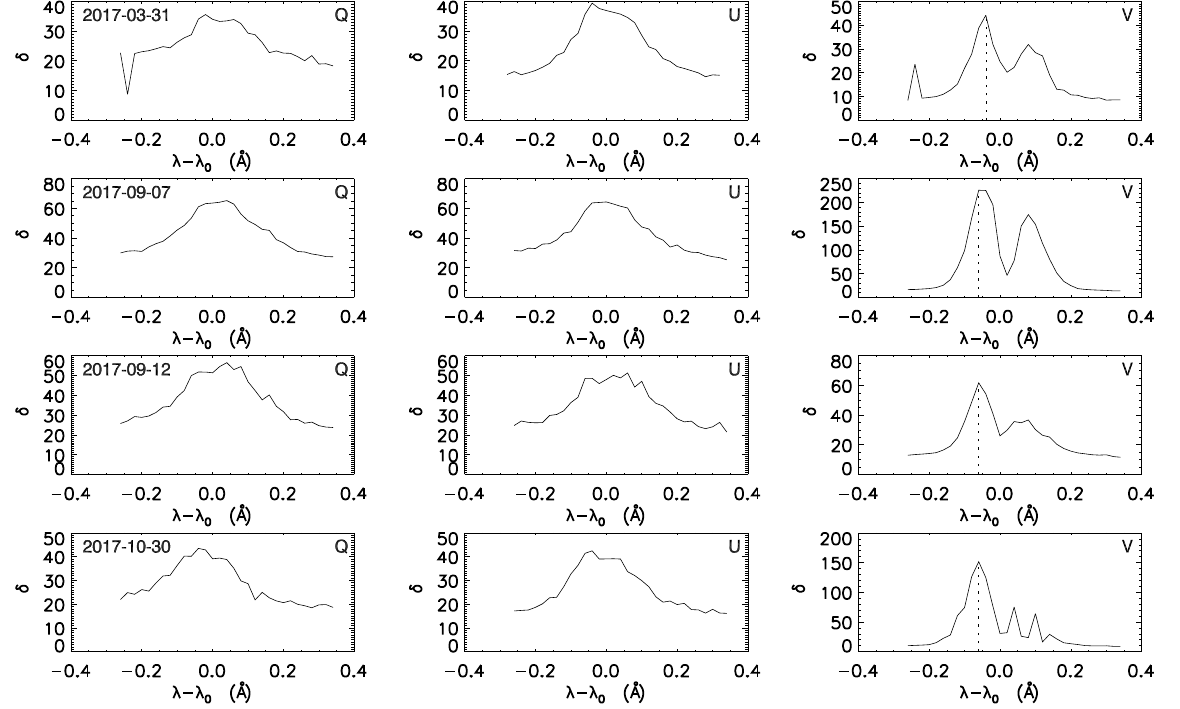}}
\caption{Variations in magnetic sensitivity for Stokes-Q, U, and V within a quiet region (marked by a white square in Figure \ref{31q}) over four days. The horizontal axis represents the wavelength offset from the line center ($\lambda-\lambda_{0}$) in \AA), while the vertical dotted lines in the third column pinpoint the peak variance ($\delta$s) values for each parameter.} \label{sigmafor4days}
\end{figure}

Figure \ref{sigmafor4days} delves deeper into the nuances of magnetic sensitivity, showcasing the variations in variance ($\delta$) for Stokes-Q, U, and V within a quiet region at the disk center (as seen in Figure \ref{31q}) over four days. The horizontal axis represents the observed wavelength positions relative to the line center, spanning from -0.3~\AA~to 0.3~\AA. Notably, the vertical dotted lines in the third column pinpoint the peak $\delta$ values for Stokes-V, ranging from -0.06 and -0.08\, \AA~with a median of -0.07\,\AA. These results echo the findings of Figure \ref{31q}: Stokes-Q and U exhibit peak sensitivity at the line center, while Stokes-V reaches its maximum sensitivity slightly blueward of the center.

\newpage
\section{Theoretical Analysis}\label{sec:Theoretical Analysis}

Building upon the Milne-Eddington (M-E) atmosphere model, we leverage analytical solutions for the radiative transfer equations governing polarized light (\citeauthor{1956PASJ....8..108U} \citeyear{1956PASJ....8..108U}; \citeauthor{1967IzKry..37...56R} \citeyear{1967IzKry..37...56R}; \citeauthor{1982SoPh...78..355L} \citeyear{1982SoPh...78..355L}). In this study, we fix specific parameters associated with these solutions: the spectral line damping parameter (a = 0.95), line center wavelength ($\lambda_{0}$=5324.186\,\AA), line center opacity ratio ($\eta_{0}$=10.47), Doppler width ($\Delta\lambda_{D}$=40.50), and source function ($\mu B_{1}=-1.16$). However, the magnetic field parameters - total field strength [$B$], inclination [$\gamma$] and azimuth [$\phi$] - remain variable.

Response functions quantify the sensitivity of Stokes parameter profiles to changes in atmospheric properties. They are formulated as partial derivatives of the Stokes vector with respect to the modeled parameters (\citeauthor{1994A&A...283..129R} \citeyear{1994A&A...283..129R}; \citeauthor{2007A&A...462.1137O} \citeyear{2007A&A...462.1137O}). Mathematically, they take the form:

\begin{equation}
R_{x}(\lambda)=\partial I(\lambda)/\partial x.
\end{equation}
where $R_{x}(\lambda)$ is a function of wavelength and $x$ represents any parameter within the analytical solutions. Here, we focus on response functions relating to magnetic field strength and utilize these results as valuable guidance for FMG to select an optimal wavelength position for routine observations.

\begin{figure}

\centerline{\includegraphics[width=1\textwidth,clip=]{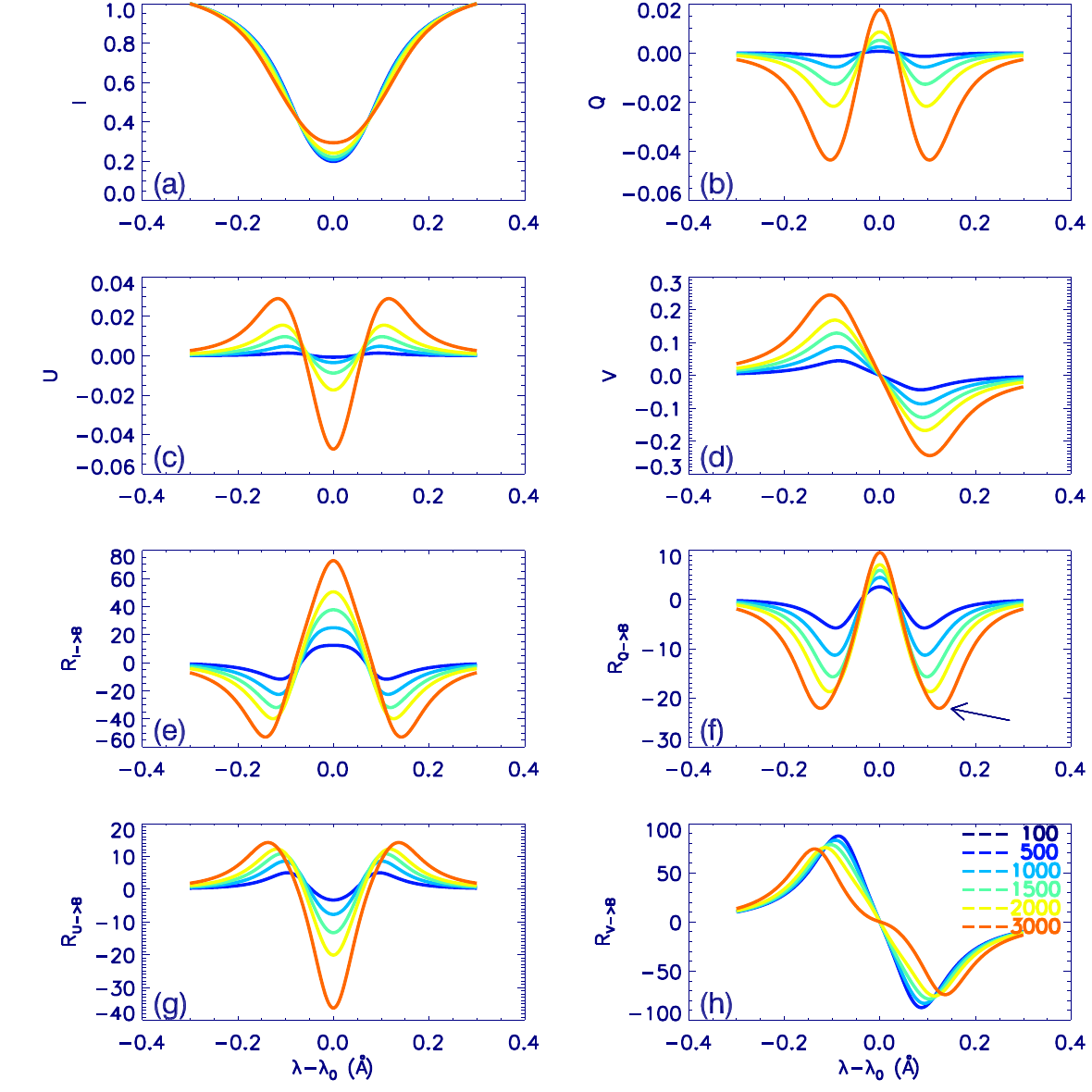}}
\caption{Panels \textbf{(a)}, \textbf{(b)}, \textbf{(c)}, and \textbf{(d)} individually show I, Q, U and V profiles and Panels \textbf{(e)}, \textbf{(f)}, \textbf{(g)}, and \textbf{(h)} give their response to field strength, where the thermodynamic parameters are fixed with a=0.95, $\lambda_{0}$=5324.186\AA, $\eta_{0}$=10.47, $\Delta\lambda_{D}$=0.04050\AA, $\mu B_{1}=-1.16$, and inclination ($\gamma$) and azimuth ($\phi$) angls are assigned 45.0$^{\circ}$ and 67.5$^{\circ}$, while magnetic field strength (B) be changed from the values of 20, 50, 100, 200, 500, 1000, 1500, 2000, 2500, to 3000 (G). Where an arrow in Panel \textbf{(f)} indicates the local peak position of spectral line in response fuction.} \label{quv_rf_rvb_prfile}
\end{figure}

\begin{figure}
\centerline{\includegraphics[width=1\textwidth,clip=]{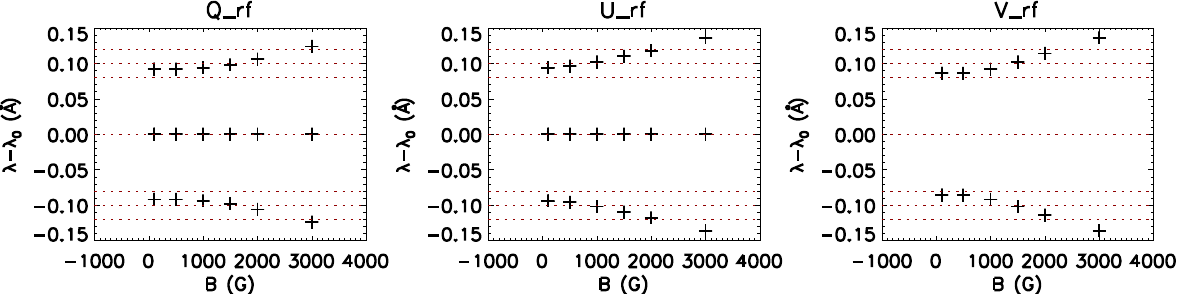}}
\caption{Local peak of the response functions vs. field strength extracted from Figure \ref{quv_rf_rvb_prfile}. The horizontal lines label $\lambda-\lambda_{0}$=0, $\pm$0.08, $\pm$0.1, $\pm$0.12 (\AA).} \label{quv_rf_rvb}
\end{figure}

Figure \ref{quv_rf_rvb_prfile} shows the responses of I, Q, U and V to B with the thermodynamic parameters, inclination ($\gamma$) and azimuth ($\phi$) angles fixed, and the magnetic field strength (B) assigned with the values of 100, 500, 1000, 1500, 2000, 3000 (G). In this Figure, the panels a - d show the profiles of I, Q, U and V under different magnetic field strengths, while the panels e - h show the corresponding response functions, respectively. It can be found that the more sensitive positions for Q and U are located at line center, while those for V located at the wings, where their values change significantly with wavelength. Additionally, with the field strength increasing the sensitive positions of V shift toward line wings. For Q and U, there are local peak values at line wings as indicated by an arrow in the panel f, which also shift to line wings with field strength increasing.
There are better sensitivities at the spectral position with local peak than those with monotonous curves in response function.
To show the results more intuitively, the local peak values among these response functions in Figure \ref{quv_rf_rvb_prfile} are extracted out and their variations with field strength are shown in Figure \ref{quv_rf_rvb}. In the figure, the horizontal lines with values of $\lambda-\lambda_{0}$=0, $\pm$0.08, $\pm$0.1, $\pm$0.12\,\AA~are plotted. It can be found that below 3000 Gauss, for Q and U there are two spectral bands from 0.08 to 0.15\,\AA, which basically match with that for V.

\section{Analysis of the in-orbit Test Data of FMG}
\label{sect:The FMG Test}
\begin{figure}
\centerline{\includegraphics[width=1\textwidth,clip=]{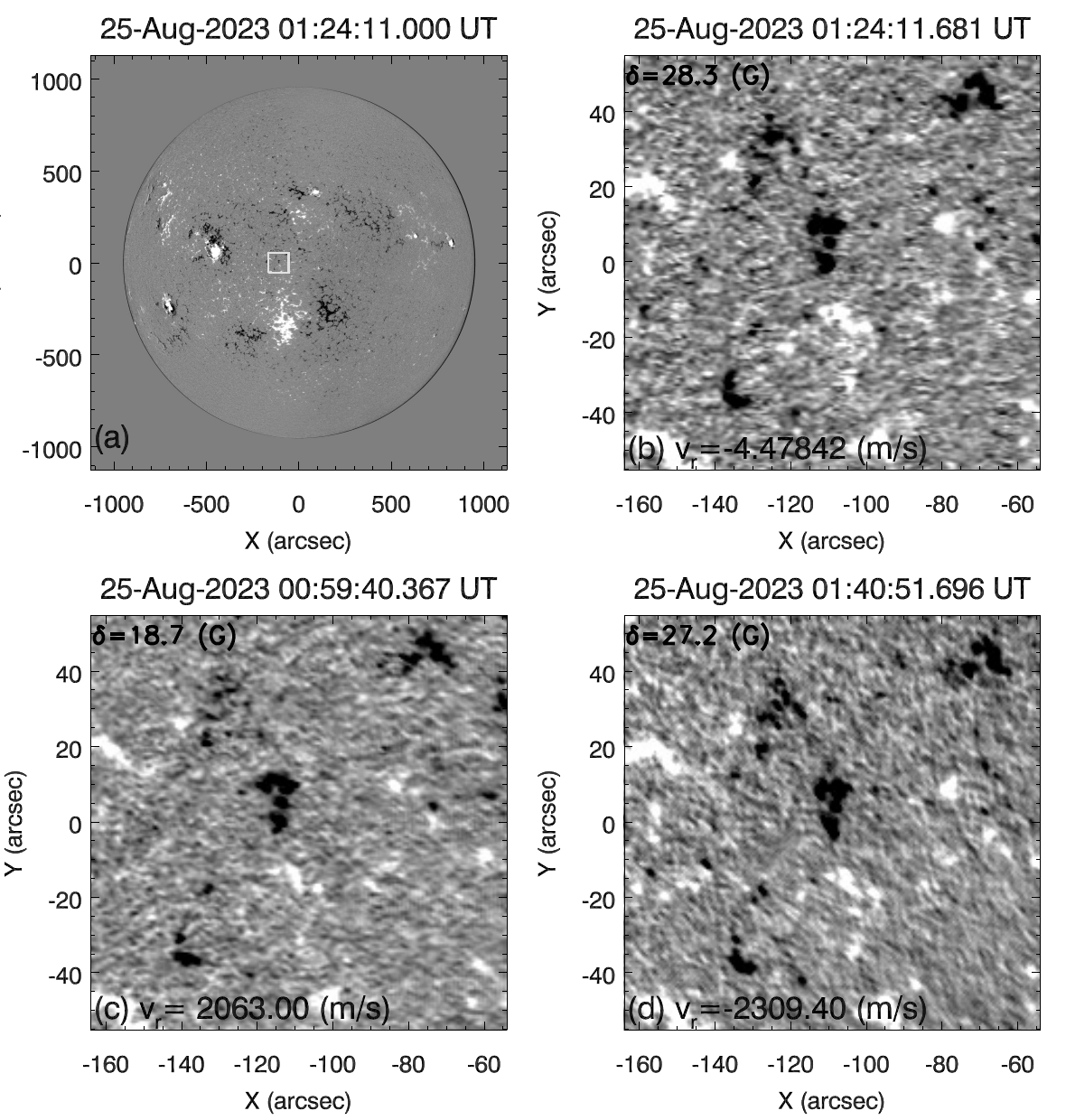}}
\caption{Panel \textbf{(a)} shows a full-disk LOS magnetic field observed on 25 Agust 2023, the \textit{white squre} identifies a quiet region selected for further investigation. Panels \textbf{(b)}, \textbf{(c)}, and \textbf{(d)} show the LOS magnetic field of within this region at three distinct Sun-FMG radial velocities ($v_r$): -4.4 m/s ($\delta$ = 28.3 G), 2063.0 m/s ($\delta$ = 18.7 G), and -2309.4 m/s ($\delta$ = 27.2 G), respectively.} \label{fmgdatareg}
\end{figure}

\begin{figure}
\centerline{\includegraphics[width=1\textwidth,clip=]{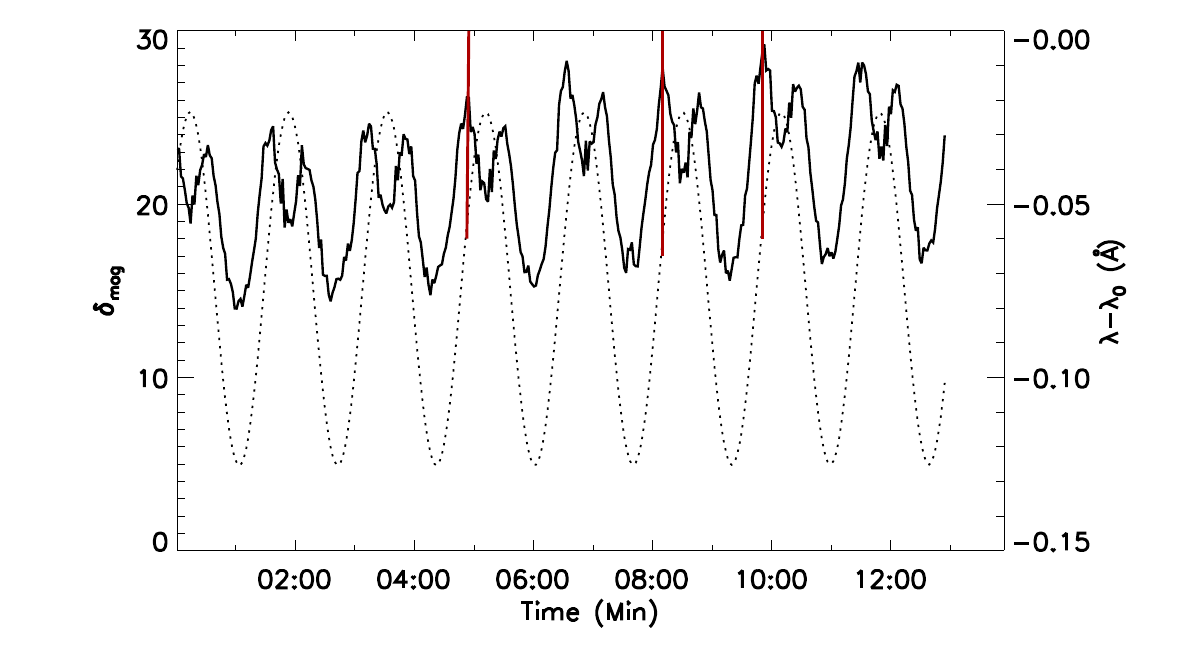}}
\caption{$\delta$s, in the quiet region as outlined by a \textit{white squre} in Figure\ref{fmgdatareg}, vs. the corresponding $\delta\lambda$=$\lambda-\lambda_0$, observed on 15 Agust 2023. Three vertical red lines indicate the local peaks of $\delta$s.} \label{fmgdataregsigvsvr}
\end{figure}

ASO-S satellite was launched with the CZ-2D rocket at 15:43:55 UT on 09 October 2022.
FMG started its first operation on 16 October 2022. It has the scanning mode for spectral calibration, however, for safety this mode is not available at present. An alternative scheme based on the velocity of satellite can be used to adjust the spectral position approximately. The ASO-S satellite operates in a sun synchronous orbit (SSO) with an altitude of about 720 km, an orbital inclination angle of $98.27\deg$, and an orbital period of about 99.2 minutes. All these together result in the Sun--FMG radial velocity [$v_{r}$] changing from -3.9 to +3.9 km\,s$^{-1}$ (the positive direction is the normal direction of the sun--the sun points towards the satellite). Correspondingly, a wavelength shift ranging from -0.07\,\AA~ to +0.07\,\AA{} will superimpose on the selected spectral position. Hence, the magnetic field can be obtained at various spectral position depending on variations of the Sun--FMG radial velocity.
In FMG routine observation, if the filter position is fixed at -0.08\,\AA~away from the line center, then the actual spectral position can be deduced with $v_{r}$. In Figure \ref{fmgdatareg}, panel a shows a full-disk line-of-sight (LOS) magnetogram observed at 01:24:11 UT on 25 August 2023, where the white square region is selected for investigating the signals of LOS magnetic field obtained at different $v_r$. Panels b, c, and d show the LOS magnetogram in the quiet region when $v_r$ = -4.4 m\,s$^{-1}$, $v_r$ =2063.0 m\,s$^{-1}$, and $v_r$ =-2309.4 m\,s$^{-1}$ observed on 25 February 2023, respectively. We assume that the changes of LOS magnetic field during this time interval mainly result from the variations of spectral position caused by $v_{r}$.
It can be found that the signal of LOS magnetic field observed at $v_{r}$=-4.4 m\,s$^{-1}$ ($\delta$=28.3) is better than those at $v_r$ =2063.0 m\,s$^{-1}$ ($\delta$=18.7) and $v_r$ =-2309.4 m\,s$^{-1}$ ($\delta$=27.2). Figure \ref{fmgdataregsigvsvr} shows variance ($\delta$s) in the quiet region as indicated by the white square in Figure \ref{fmgdatareg}, varying with the corresponding $\delta\lambda$s observed on 15 August 2023, which are deduced from the various $v_{r}$. It is found that the local peak of $\delta$s is located at around $\delta\lambda$ -0.065\,\AA~(in a range from -0.08 to -0.05\,\AA), e.g., those indicated by three red vertical lines. It suggests that these results are basically consistent with those of the SMAT experimental data as well as the theoretical analysis of radiative transfer equations.

\section{Discussion and Conclusions}\label{sec:disc-conc}

This study delved into the magnetic sensitivity of the Fe {\sc i} $\lambda$5234.19 \AA~line to varying field strengths, drawing upon a multifaceted approach: SMAT observations at HSOS, in-orbit tests of the FMG onboard ASO-S, and theoretical analysis of radiative transfer response functions. The insights gained provide valuable guidance for selecting the optimal spectral position for routine FMG observations.

Both SMAT and FMG were specifically designed for full-disk magnetic field measurements using the same Fe {\sc i} $\lambda$5234.19 \AA~line, making SMAT an ideal candidate for this investigation. Our analysis revealed that the most sensitive positions differ for different Stokes parameters: Q and U exhibit peak sensitivity at the line center, while V reaches its maximum about -0.07\,\AA~away from the center. It is important to note, however, that our study focused on a quiet region at the disk center with relatively low field strengths.

The theoretical investigation, employing analytical solutions and their corresponding response functions, corroborated the findings from the SMAT data. Both approaches identified the line center as the most sensitive position for Q and U, while V demonstrated peak sensitivity within the line wing region. Interestingly, for field strengths below 3000 G, both Q/U and V exhibit shared zones (bands) of increased sensitivity around 0.08-0.15 \AA. This highlights the advantage of choosing a common spectral position for FMG observations of all three Stokes parameters. Additionally, within these bands, the peak sensitivity for both Q/U and V tends to shift towards the line wings as the magnetic field strength increases.
Thus, it can be inferred that different spectral line positions should be used to measure different magnetic field intensities. Therefore, when paying attention to magnetic field of different features or structure of active region, the corresponding spectral line positions should be selected for specific observation targets. Consquently, it is a difficult task to choose spectral position to observe magnetic field, when the whole active region regards as an observation target. It is difficult to simultaneously balance the weak magnetic field (e.g. $<$100 G) in the quiet area and the strong magnetic field (e.g. $>$2000 G) in the umbra area for a fixed wavelength position. In practical observations, it is crucial to meticulously and comprehensively consider the determination of observation spectral line positions in order to measure the magnetic field and generate high-quality data with minimal errors. For instance, when an active region is considered as an object with an expected magnetic field strength ranging from 500 to 1500 G, a magnetic field intensity of 1000 G can be utilized as a benchmark for selecting the appropriate spectral position, which indicates that the entire magnetic field is measured using the optimal spectral position, resulting in the minimum error in magnetic field measurement.

Optimizing the selection of magnetically sensitive spectral line positions is paramount for precise observations of magnetic fields. However, calibrating the magnetic field is equally crucial to ensure accurate measurements. Under the weak field approximation and application of linear calibration, the varying strength of the magnetic field across distinct solar features leads to varying error amplitudes (\citeauthor{2019SCPMA..6299601Z}, \citeyear{2019SCPMA..6299601Z}). \citeauthor{2019SCPMA..6299601Z} (\citeyear{2019SCPMA..6299601Z}) studied the relationship between Stokes V/I and magnetic field strength across various spectral positions. The results showed that the curve shapes of the V/I-magnetic field strength relationship vary depending on the spectral position. Therefore, even for a magnetic field with a fixed strength, the errors in determining the magnetic field under the weak field approximation would differ when different spectral positions are used. For instance, if we consider a fixed magnetic field of 1500 G, the errors in magnetic field determination using linear calibration under the weak field approximation would be 84 G and 72 G for the spectral lines selected at -0.035\,\AA~ and -0.075\,\AA, respectively. It should be noted that the magnetic field error also depends on the linearity degree of the magnetic field and V/I at the fixed spectral position when linear calibration is used.

From the relations between variances ($\delta$s) of FMG LOS magnetic field versus the radial velocity of ASO-S relative to Sun ($v_r$), it can be found that the most sensitive position of Stokes V is located at the line wing of about -0.065\,\AA. It should be noted that it is symmetry for response function in theory, which means the magnetic-sensitivities are the same at the symmetrical positions of the red-shift and blue-shift. However, in observation for the SMAT and FMG the most magnetic-sensitive positions tend to locate at the blue wing of spectral line, which can be seen from the second local peaks (first local peaks in red wing) of the curves in the third column in Figure \ref{sigmafor4days}
, where variances ($\delta$s) in red-shift are lower than those in the corresponding blue-shift symmetrical position. The results of FMG test confirm the conclusions of SMAT and theoretical analysis for Stokes-V (LOS magnetic field related), and the spectral line position that is set at $-0.08\,\AA$ away from the line center, should be a suitable selection for magnetic field measurement by FMG.

\begin{acks}
This work is supported by National Key R\&D Program of China (Nos. 2022YFF0503001, 2021YFA1600500 and 2021YFA1600503), the Strategic Priority Research Program on Space Science of Chinese Academy of Sciences (Grant No. XDB0560000), and Natural Science Foundation of China (Grant Nos. 12273059, 11203036, 11703042, U1731241, 11427901, 11473039, 11427901 and 11178016), and
the Youth Innovation Promotion Association of CAS(2023061).
ASO-S mission is supported by the Strategic Priority Research Program on Space Science, the Chinese Academy of Sciences, Grant No. XDA15320000.
\end{acks}

\begin{authorcontribution}
\end{authorcontribution}

\begin{fundinginformation}
\end{fundinginformation}

\begin{dataavailability}
\end{dataavailability}

\begin{ethics}
\begin{conflict}
\end{conflict}
\end{ethics}

\bibliographystyle{spr-mp-sola}
\bibliography{liu}

\end{document}